\RequirePackage[mathlines]{lineno}
\documentclass[prd,twocolumn,amsmath,amssymb]{revtex4}
\usepackage{overpic,graphicx}
\usepackage{dcolumn}
\usepackage{bm}
\usepackage{rotating}
\usepackage{subfigure}
\usepackage{color}
\usepackage{multirow}
\usepackage{epstopdf}
\usepackage{mathrsfs}
\usepackage{comment}
\usepackage{booktabs}
\usepackage{defs_LamcPEtap}
\usepackage{gensymb}
\usepackage{hyperref}
\usepackage{diagbox}

\renewcommand\tablename{\rm Table}

\begin{document}

\title{\bf \boldmath
Measurement of the absolute branching fraction of the singly Cabibbo suppressed decay $\Lambda^{+}_{c}\to p\eta^{\prime}$
}

\author{
\begin{small}
\begin{center}
M.~Ablikim$^{1}$, M.~N.~Achasov$^{11,b}$, P.~Adlarson$^{70}$, M.~Albrecht$^{4}$, R.~Aliberti$^{31}$, A.~Amoroso$^{69A,69C}$, M.~R.~An$^{35}$, Q.~An$^{66,53}$, X.~H.~Bai$^{61}$, Y.~Bai$^{52}$
% ...
\\
\vspace{0.2cm}
(BESIII Collaboration)\\
\vspace{0.2cm} {\it
$^{1}$ Institute of High Energy Physics, Beijing 100049, People's Republic of China\\
% ...
$^{4}$ Bochum Ruhr-University, D-44780 Bochum, Germany\\
% ...
$^{11}$ G.I. Budker Institute of Nuclear Physics SB RAS (BINP), Novosibirsk 630090, Russia\\
% ...
$^{31}$ Johannes Gutenberg University of Mainz, Johann-Joachim-Becher-Weg 45, D-55099 Mainz, Germany\\
% ...
$^{35}$ Liaoning Normal University, Dalian 116029, People's Republic of China\\
% ...
$^{52}$ Southeast University, Nanjing 211100, People's Republic of China\\
$^{53}$ State Key Laboratory of Particle Detection and Electronics, Beijing 100049, Hefei 230026, People's Republic of China\\
% ...
$^{61}$ University of Jinan, Jinan 250022, People's Republic of China\\
% ...
$^{66}$ University of Science and Technology of China, Hefei 230026, People's Republic of China\\
% ...
$^{69}$ University of Turin and INFN, (A)University of Turin, I-10125, Turin, Italy; (B)University of Eastern Piedmont, I-15121, Alessandria, Italy; (C)INFN, I-10125, Turin, Italy\\
$^{70}$ Uppsala University, Box 516, SE-75120 Uppsala, Sweden\\
% ...
$^{b}$ Also at the Novosibirsk State University, Novosibirsk, 630090, Russia\\
% ...
}
\end{center}
\vspace{0.4cm}
\end{small}
}

\begin{abstract}
The singly Cabibbo suppressed decay $\LamCPE$ is measured using 4.5
$\invfb$ of $\ee$ collision data collected at center-of-mass energies
between 4.600 and 4.699 GeV with the BESIII detector at BEPCII.  Evidence
for $\LamCPE$ with a statistical significance of $3.6\sigma$ is
reported with a double-tag approach.  The $\LamCPE$ absolute
branching fraction is determined to be $(5.62^{+2.46}_{-2.04} \pm
0.26)\times 10^{-4}$, where the first and second uncertainties are
statistical and systematic, respectively.  Our result is consistent with the
branching fraction obtained by the Belle collaboration within the uncertainty of $1\sigma$.
\end{abstract}

\maketitle

\oddsidemargin  -0.2cm
\evensidemargin -0.2cm

\section{Introduction}

The weak decays of the ground state charmed baryon $\LamC$ play an
essential role in understanding the interplay of weak and strong
interactions in the charm region~\cite{Cheng:2015iom}.  In addition,
information about the lightest charmed baryon provides key input for
investigations of heavier charmed baryons~\cite{Sheng:2018dc} and
bottom baryons~\cite{Rupak:2016bb, Detmold:2015bb}.  In contrast to
charmed meson decays, which are usually dominated by factorizable
amplitudes, decays of charmed baryons receive sizable non-factorizable
contributions which arise from internal $W$-emission and
$W$-exchange~\cite{Chau:1986bb,Kohara:1991bb}.

The complicated physics in charmed baryon decays is described by
phenomenological model calculations, which strongly rely on
experimental results.  Experimentally, some Cabibbo favored
decays~\cite{Zupanc:2013iki, Ablikim:2015flg, Ablikim:2016mcr,
  Belle:2017tfw, Ablikim:2018czr, Ablikim:2020ffk} have been measured
with relatively high precision, while the singly Cabibbo suppressed
(SCS) decays are limited by statistics. Recently, measurements of SCS
decay branching fractions have been carried out by the BESIII and
Belle Collaborations~\cite{Ablikim:2016phi, Ablikim:2017eta,
  Belle:2021pi0}.  The SCS decay $\Lambda_c^+ \to n\pi^{+}$ was
observed for the first time at BESIII~\cite{Ablikim:2022SCS}, and the
measured branching fraction is consistent with the SU(3)
prediction~\cite{Geng:2018rse} but twice as large as that from current
algebra~\cite{Cheng:2018hwl}.

\begin{figure}[!htp] \begin{center}
\subfigure[]{\includegraphics[width=0.23\textwidth]{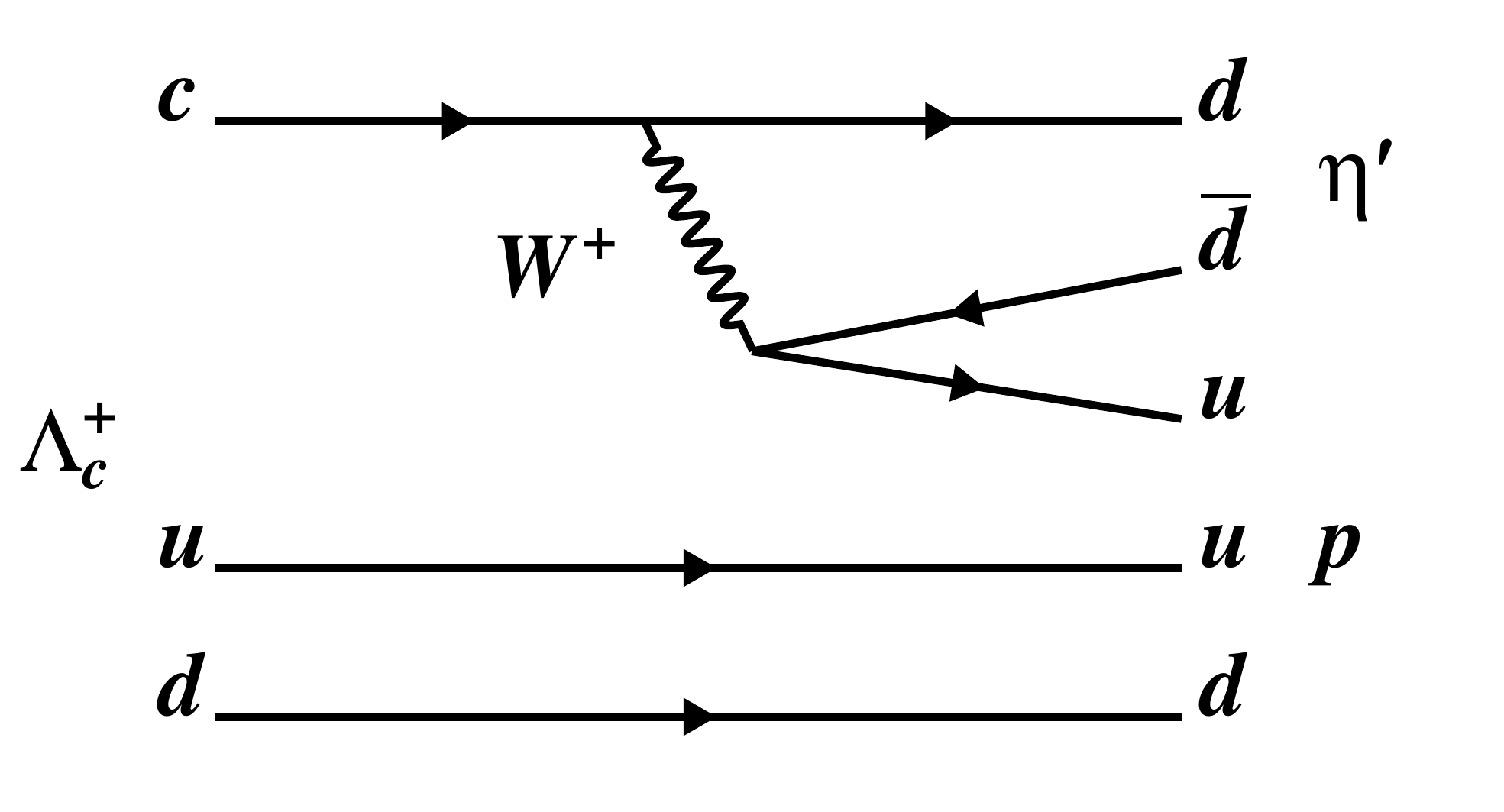}\label{fig:feynman_emission}}
\subfigure[]{\includegraphics[width=0.23\textwidth]{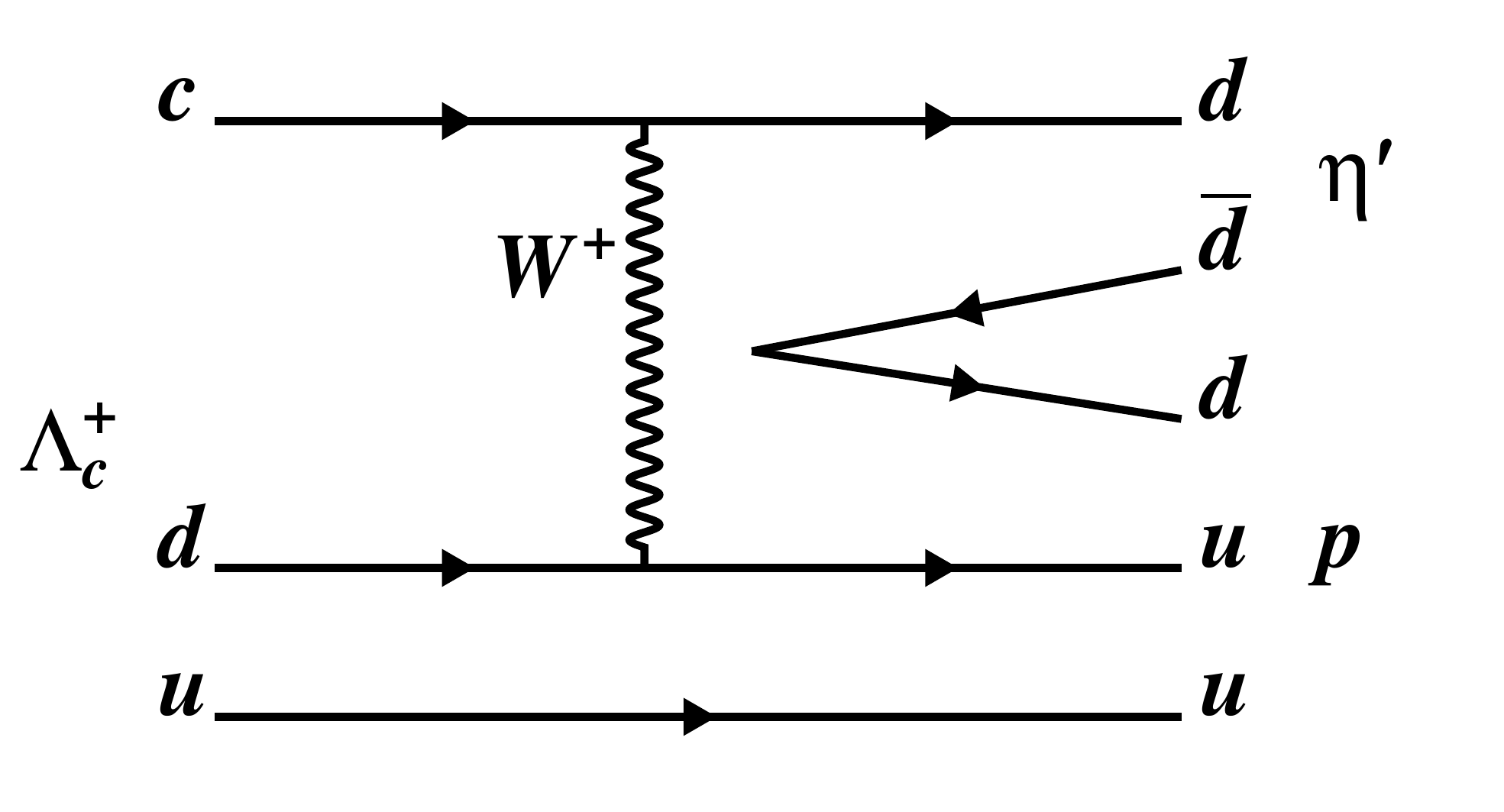}\label{fig:feynman_exchange}}
\end{center} \caption{ Lowest-order Feynman diagrams for $\LamCPE$
mediated via (a) $W$-emission and (b) $W$-exchange mechanisms.
      } \label{fig:feynman_petap}
\end{figure}

The Belle Collaboration reported the first observation of
the SCS decay $\LamCPE$ with a statistical significance of 5.4$\sigma$
and measured its decay branching fraction with respect to the
$\Lambda_c^{+}\to pK^-\pi^+$ to be
$\mathcal{B}(\LamCPE)/\mathcal{B}(\Lambda_c^{+}\to pK^-\pi^+)= (7.54
\pm 1.32 \pm 0.75)\times 10^{-3}$~\cite{Belle:2022etap}. The two-body
SCS decay $\LamCPE$ can proceed via the internal $W$-emission and
$W$-exchange mechanisms, with the lowest-order Feynman diagrams shown
in Fig.~\ref{fig:feynman_petap}.  The branching fraction of $\LamCPE$
is predicted to be in the range of $10^{-3} -
10^{-4}$~\cite{Cheng:2018hwl}.   To improve the knowledge of charmed
baryons, more experimental studies of SCS decays are highly
desirable.

In this paper we report a measurement of the absolute branching
fraction of the SCS decay $\LamCPE$ using 4.5 $\invfb$ of $\ee$
collision data collected with the BESIII detector at seven
center-of-mass (CM) energies between 4.600 and 4.699 GeV.  The $\LCpair$
data sets are above $\LCpair$ threshold and provide a clean
environment with which to measure the absolute branching fractions of
$\Lambda^+_c$. The CM energy and the integrated luminosity for each
energy point are listed in
\tablename~\ref{tab:data_sets}~\cite{BESIII:lumi0, BESIII:lumi1,
BESIII:lumi2}.  Throughout the text, the charge conjugate states are
always implied.

\begin{table}[!htbp]
  \begin{center}
    \caption{The CM energy and the integrated luminosity ($\mathcal{L}_{\rm int}$) 
         for each energy point.
             The first and the second uncertainties are statistical and systematic, respectively.}
    \begin{tabular}{ c  c  c }

      \hline
      \hline
         Data set  &   CM energy (MeV) &  $\mathcal L_{\rm int}$ (\ipb)  \\
      \hline

            4600  &   4599.53 $\pm$ 0.07 $\pm$ 0.74   &~~  586.90  $\pm$ 0.10 $\pm$ 3.90 \\
            4612  &   4611.86 $\pm$ 0.12 $\pm$ 0.32   &~~  103.83  $\pm$ 0.05 $\pm$ 0.55 \\
            4628  &   4628.00 $\pm$ 0.06 $\pm$ 0.32   &~~  521.52  $\pm$ 0.11 $\pm$ 2.76 \\
            4641  &   4640.91 $\pm$ 0.06 $\pm$ 0.38   &~~  552.41  $\pm$ 0.12 $\pm$ 2.93 \\
            4661  &   4661.24 $\pm$ 0.06 $\pm$ 0.29   &~~  529.63  $\pm$ 0.12 $\pm$ 2.81 \\
            4682  &   4681.92 $\pm$ 0.08 $\pm$ 0.29   &~   1669.31 $\pm$ 0.21 $\pm$ 8.85 \\
            4699  &   4698.82 $\pm$ 0.10 $\pm$ 0.39   &~~  536.45  $\pm$ 0.12 $\pm$ 2.84 \\

      \hline\hline
    \end{tabular}
      \label{tab:data_sets}
  \end{center}
\end{table}

\section{BESIII Detector and Monte Carlo simulation}
The BESIII detector~\cite{Ablikim:2009aa} records
symmetric $e^+e^-$ collisions provided by the BEPCII storage
ring~\cite{Yu:IPAC2016-TUYA01} in the CM energy range from 2.0 to
4.95~GeV~\cite{Ablikim:2019hff}, with a peak luminosity of $1 \times
10^{33}\;\text{cm}^{-2}\text{s}^{-1}$ achieved at CM energy of
$3.77\;\text{GeV}$.  BESIII has collected large data samples in this
energy region~\cite{Ablikim:2019hff}.  The cylindrical core of the
BESIII detector covers 93\% of the full solid angle and consists of a
multilayer drift chamber~(MDC) operating with a helium-based gas
mixture, a plastic scintillator time-of-flight system~(TOF), and a
CsI(Tl) electromagnetic calorimeter~(EMC), which are all enclosed in a
superconducting solenoidal magnet providing a {\spaceskip=0.2em\relax
1.0 T} of magnetic field.  The solenoid is supported by an octagonal
flux-return yoke with resistive plate counter muon identification
modules interleaved with steel.  The charged-particle momentum
resolution at $1~{\rm GeV}/c$ is $0.5\%$, and the resolution of the
ionization energy loss in the MDC ($\mathrm{d}E/\mathrm{d}x$) is $6\%$
for electrons from Bhabha scattering. The EMC measures photon energies
with a resolution of $2.5\%$ ($5\%$) at $1$~GeV in the barrel (end
cap) region. The time resolution in the TOF barrel region is 68
ps. The end cap TOF system was upgraded in 2015 using multi-gap
resistive plate chamber technology, providing a time resolution of 60
ps~\cite{LiGuo:MRPC}.

Simulated data samples are produced with {\sc
geant4}-based~\cite{Agostinelli:2002hh} Monte Carlo (MC) software,
which describes the geometry of the BESIII detector and simulates the
performance of the detector.  The signal MC samples of $\ee\to\LCpair$
with $\LamCB$ decaying into ten specific tag modes (as described below
and listed in \tablename~\ref{tab:yield-st-468}) and $\LamCPE$, which
are used to determine the detection efficiencies, are generated for
each individual CM energy using the generator {\sc
kkmc}~\cite{Jadach:2000ir} incorporating initial-state radiation
(ISR) effects and the beam energy spread.  The $\LamCPE$ decay is
modeled with a uniform phase-space distribution.  The inclusive MC
samples, which consist of $\LCpair$ events, charmed meson
$D_{(s)}^{(\ast)}$ pair production, ISR return to the
charmonium(-like) $\psi$ states at lower masses, and continuum
processes $e^{+}e^{-}\rightarrow q\bar{q}$ ($q=u,d,s$), are generated
to estimate the potential background. Particle decays are modeled with
{\sc evtgen}~\cite{Lange:2001uf, Ping:2008zz} using branching
fractions taken from the Particle Data Group
(PDG)~\cite{PDG:2020}, when available, or otherwise estimated with
{\sc lundcharm}~\cite{Chen:2000tv, YANG:2014}.  Final state radiation
from charged final state particles is incorporated using {\sc
photos}~\cite{Richter-Was:1992hxq}.

\section{METHODOLOGY}

A double-tag (DT) approach~\cite{MarkIII:DT} is implemented to study
the SCS decay $\LamCPE$. A data sample of $\LamCB$ baryons, referred to
as the single-tag (ST) sample, is reconstructed with ten exclusive
hadronic decay modes, as listed in \tablename~\ref{tab:yield-st-468}.
Events in which the signal decay $\LamCPE$ is reconstructed in the
system recoiling against the $\LamCB$ candidates of the ST sample are
denoted as DT candidates.  The branching fraction of $\LamCPE$ is
determined as

\begin{equation} \label{eq:br}
  \mathcal{B}(\LamCPE)=\frac{N_{\mathrm{sig}}} {\br_{\rm inter} \cdot
  \sum_{ij} N_{ij}^{\mathrm{ST}}\cdot
  (\epsilon_{ij}^{\mathrm{DT}}/\epsilon_{ij}^{\mathrm{ST}}) },
  \end{equation} where $N_{\mathrm{sig}}$ is the signal yield of the
  DT candidates and $\br_{\rm inter}$ is the $\eta^{\prime}$ decay
  branching fraction taken from the PDG~\cite{PDG:2020}.  The
  subscripts $i$ and $j$ represent the ST modes and the data samples
  at different CM energies, respectively. The parameters
  $N_{ij}^{\mathrm{ST}}$, $\epsilon_{ij}^{\mathrm{ST}}$ and
  $\epsilon_{ij}^{\mathrm{DT}}$ are the ST yields, ST and DT detection
  efficiencies, respectively.

\section{Single Tag Event Selections} \label{sec:single-tag}

The selection criteria of ST events are same as the Ref.~\cite{Ablikim:2022SCS}.
Charged tracks detected in the MDC are required to be within a polar
angle ($\theta$) range of $|\!\cos\theta| < 0.93$, where $\theta$ is
defined with respect to the $z$-axis, which is the symmetry axis of
the MDC.  Except for tracks from $\Ks$ and $\bar{\Lambda}$ decays,
their distances of closest approach to the interaction point (IP) must
be less than 10~cm along the $z$-axis, and less than 1~cm in the
transverse plane (referred to as a tight track hereafter).
Particle identification (PID) is implemented by combining the
measurements of $\mathrm{d}E/\mathrm{d}x$ in the MDC and the
flight time in the TOF, and each charged track is assigned a particle
type of pion, kaon or proton, according to which assignment has the
highest probability.

Photon candidates are identified from showers in the EMC. The deposited
energy of each shower must be more than 25~MeV in the barrel region
($|\!\cos\theta| \le 0.80$) or more than 50~MeV in the end cap region
($0.86 \le |\!\cos\theta| \le 0.92$). To suppress electronic noise and
showers unrelated to the event, the difference between the EMC time
and the event start time is required to be within [0, 700]~ns. A
$\pi^0$ candidate is reconstructed with a photon pair within the invariant
mass region (0.115, 0.150)~GeV/$c^2$. To improve the resolution, a
kinematic fit is performed by constraining the invariant mass of the
photon pair to the world average $\pi^0$ mass~\cite{PDG:2020}.  The
momentum updated by the kinematic fit is used in further analysis.

Candidates for $\Ks$ and $\bar{\Lambda}$ are reconstructed in their
decays to $\pi^+\pi^-$ and $\bar{p}\pi^+$, respectively, where the
charged tracks must have the distances of closest approach to the IP
within 20~cm along the $z$-axis (referred to as a loose track
hereafter).  To improve the signal purity, PID is required for the
anti-proton candidate, but not for the charged pion.  The two daughter
tracks are constrained to originate from a common decay vertex, and
the $\chi^2$ of the vertex fit is required to be less than
100. Furthermore, the decay vertex is required to be separated from
the IP by a distance of at least twice the fitted vertex
resolution. The momenta of the $\Ks$ or $\bar{\Lambda}$ candidates updated
by the fit are used in further analysis, and the invariant masses are
required to be within (0.487, 0.511)~GeV/$c^2$ for $\pi^+\pi^-$ and
(1.111, 1.121)~GeV/$c^2$ for $\bar{p}\pi^+$.  The $\bar{\Sigma}^0$ and
$\bar{\Sigma}^-$ candidates are reconstructed with the
$\gamma\bar{\Lambda}$ and $\bar{p}\pi^0$ final states with invariant
masses being within (1.179, 1.203) and (1.176, 1.200)~GeV/$c^2$,
respectively.

The ST $\LamCB$ candidates are identified using the beam constrained
mass $M_\mathrm{BC} = \sqrt{\Ebeam^2/c^4 - | \pALC |^2/c^2}$ and
energy difference $\dE = E_{\ALamC} - \Ebeam $, where $\Ebeam$ is the
beam energy and $E_{\ALamC}$ and $\pALC$ are the energy and momentum
of the $\LamCB$ candidate, respectively.  The $\LamCB$ candidates are
required to satisfy tag-mode dependent asymmetric $\dE$ requirements,
listed in \tablename~\ref{tab:yield-st-468}, which take into account
  the effects of ISR and correspond to three times the resolution
  around the peak.  If there is more than one candidate satisfying the
  above requirements for a specific tag mode, the one with the minimum
  $|\dE|$ is kept, and those with $\mBC \in (2.275, 2.310)$~GeV/$c^2$
  are retained for further analysis.

\begin{table}[!htbp]
  \begin{center}
  \caption{Branching fractions ($\mathcal{B}_{\mathrm{ST}}$), $\dE$ requirement, ST yield, 
                and ST detection efficiency of $\LamCPE$
           of each tag mode for the data set 4682. The branching fractions have taken into account
           the intermediate particle decays, {\it i.e.} $\bar{\Lambda}\to \bar{p}\pi^+$, 
           $\bar{\Sigma}^0\to \gamma\bar{\Lambda}$, $\bar{\Sigma}^-\to \bar{p}\pi^0$,
           $\Ks\to \pi^+\pi^-$ and $\pi^0\to\gamma\gamma$.
           The uncertainty in the ST yield is statistical only. }
    \renewcommand\arraystretch{1.2}
    \begin{tabular}{ p{1.75cm} c c p{1cm}<{\raggedleft} @{ $\pm$ } p{0.75cm}<{\raggedright} c }

      \hline
      \hline
            Tag mode  & $\mathcal{B}_{\mathrm{ST}}$(\%) & $\dE$ (MeV) & \multicolumn{2}{c}{$N_{i}^{\mathrm{ST}}$}  & $\epsilon_{i}^{\mathrm{ST}}$(\%)    \\
      \hline

            $\Bpkpi$                       & 6.28 & $(-34,~20)$    &  $17,415$  & 145  & 47.3   \\
            $\Bpks$                        & 1.10 & $(-20,~20)$    &  $3,353$ & 61  &  48.1   \\
            $\bar{\Lambda}\pi^-$        &  0.83 & $(-20,~20)$    &  $2,012$ & 47   &  37.8   \\
            $\Bpkpi\pi^0$                  & 4.41 & $(-30,~20)$    &  $4,005$ & 95  & 14.5   \\
            $\Bpks\pi^0$                   & 1.35 & $(-30,~20)$    &  $1,454$ & 52    &  16.5   \\
            $\bar{\Lambda}\pi^-\pi^0$    & 4.48  & $(-30,~20)$    &  $3,576$ & 71  &  14.6   \\
            $\Bpks\pi^+\pi^-$              & 1.11 & $(-20,~20)$    &  $1,261$ & 49   &  17.7   \\
            $\bar{\Lambda}\pi^-\pi^+\pi^-$ & 2.33 & $(-20,~20)$    &   $1,818$ & 52   &  12.3   \\
            $\bar{\Sigma}^0\pi^-$          & 0.82 & $(-20,~20)$    &  $1,047$ & 34   &  19.3   \\
            $\bar{\Sigma}^-\pi^+\pi^-$     &  2.29 & $(-30,~20)$    &  $2,275$ &  63  &  16.2   \\
      \hline\hline
    \end{tabular}
    \label{tab:yield-st-468}
  \end{center}
\end{table}

For the $\LamCB\to\Bpks\pi^0$ ST mode, candidate events with $M_{\bar
p\pi^+} \in (1.100, 1.125)$~GeV/$c^2$ and $M_{\bar p\pi^0}\in
(1.170,1.200)$~GeV/$c^2$ are vetoed to avoid double counting with the
$\LamCB\to\bar{\Lambda}\pi^-\pi^0$ or
$\LamCB\to\bar{\Sigma}^-\pi^+\pi^-$ ST modes, respectively.  For the
$\LamCB\to\bar{\Sigma}^-\pi^+\pi^-$ ST mode, candidate events with
$M_{\pi^+\pi^-}\in (0.490, 0.510)$~GeV/$c^2$ and $M_{\bar p\pi^+}\in
(1.110, 1.120)$~GeV/$c^2$ are rejected to avoid double counting with
the $\LamCB\to\Bpks\pi^0$ or $\LamCB\to\bar{\Lambda}\pi^-\pi^0$ ST
modes, respectively.  In the $\LamCB\to\Bpks\pi^+\pi^-$ and
$\bar{\Lambda}\pi^-\pi^+\pi^-$ selections, candidate events with
$M_{\bar p\pi^+}\in (1.100,1.125)$~GeV/$c^2$ and $M_{\pi^+\pi^-}\in
(0.490, 0.510)$~GeV/$c^2$ are rejected, respectively.

\begin{figure}[!htp]
    \begin{center}
        \includegraphics[width=0.42\textwidth]{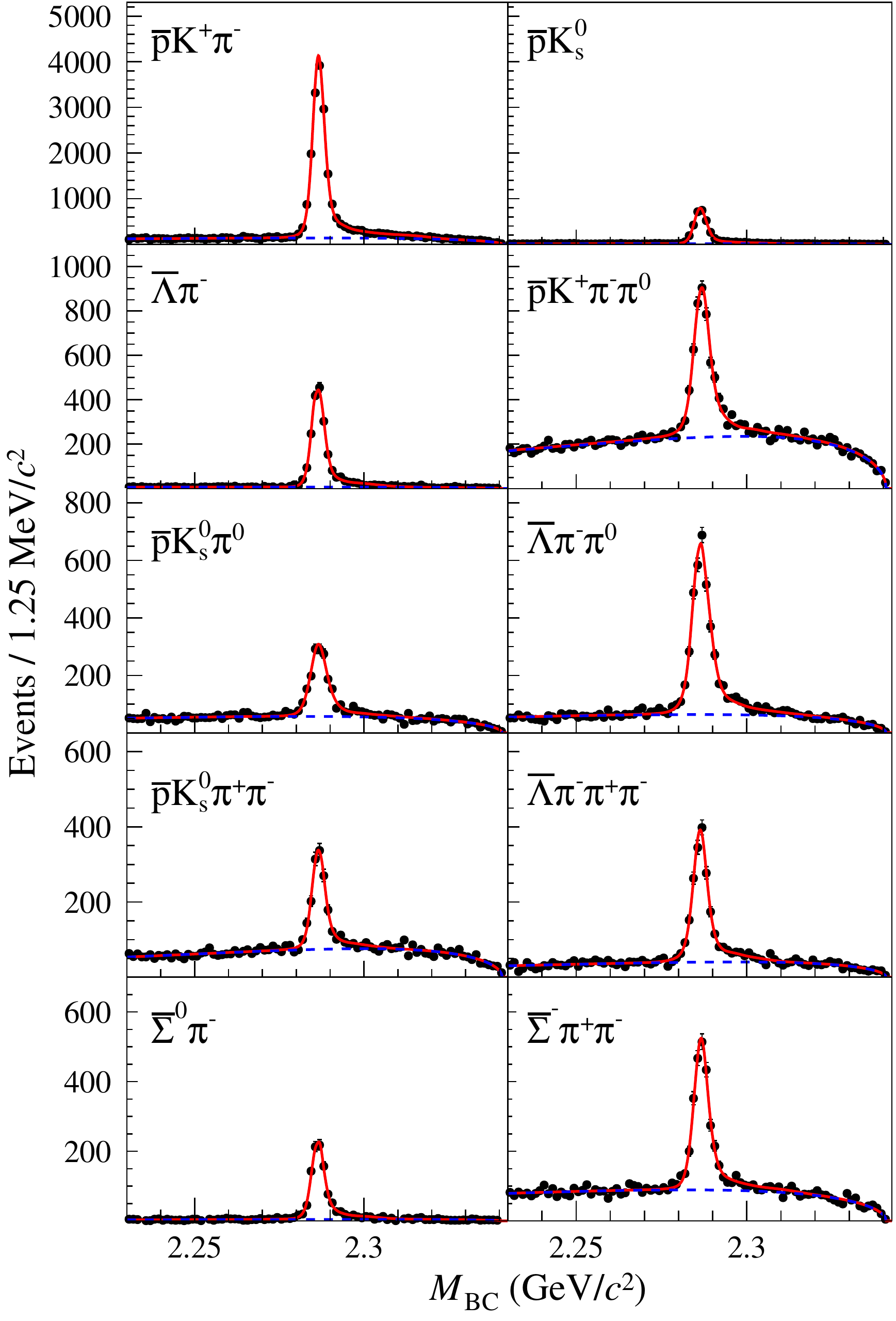}
    \end{center}
    \caption{
      The $\mBC$ distributions of the ST modes for the data set 4680.
      The points with error bars represent data. The (red) solid curves indicate the fit results and the (blue) dashed curves describe
      the background shapes.}
     \label{fig:single-tag-468}
\end{figure}

The $\mBC$ distributions of candidates for the ten ST modes with the
data set 4682 are illustrated in
Fig.~\ref{fig:single-tag-468}, where clear $\LamCB$ signals are
observed in each ST mode. No peaking background is found with the
investigation of the inclusive MC samples.  To obtain the ST yields,
unbinned maximum likelihood fits on these $\mBC$ distributions are
performed, where the signal shape is modeled with the MC-simulated
shape convolved with a Gaussian function representing the resolution
difference between data and MC simulation. The parameters of the Gaussian
function are float and determined in the fit, 
and the resultant standard deviations are varied between 0.1 and 0.7~MeV/$c^2$ for
different ST modes. 
The background shape is described by an ARGUS function~\cite{ARGUS:1990hfq}, fixing the
endpoint parameter to the corresponding $\Ebeam$. The signal yields for the individual ST modes are
summarized in \tablename~\ref{tab:yield-st-468}. The same procedure is
performed for the other six data samples at different CM energies, and the
results can be found in Ref.~\cite{Ablikim:2022SCS} and its
supplemental material.  The sum of ST yields for all data samples at
different CM energies is $(1.0524 \pm 0.0038 ) \times 10^5$, 
where the uncertainty is statistical.

\section{Signal Reconstruction}

Candidates for the signal decay $\LamCPE$ are selected with the
remaining tracks recoiling against the $\LamCB$ candidates.  The
$\eta^{\prime}$ meson is reconstructed in its two most prominent decay
modes, $\EtapEta$ and $\pi^+\pi^-\gamma$, where the $\eta$ meson is
reconstructed with its neutral decay modes, 
$i.e.$ $\eta \to 2\gamma$ and $\eta \to 3\pi^0$, 
corresponding to $(72.12
\pm 0.34)$~\%~\cite{PDG:2020} of $\eta$ decays.

For both $\eta^{\prime}$ decay modes, the signal side is required to
include exactly three tight tracks, which are then identified with PID
to be $p, \pi^{+}, \pi^{-}$.  To suppress contamination from
long-lifetime particles in the final state, the candidate events are
further required to not contain any loose tracks.  Possible background
from $\Lambda \to p\pi^-$ is rejected completely by requiring
$M_{p\pi^-} > 1.125$~GeV/$c^2$, and candidate events with
$M_{\pi^+\pi^-} \in (0.490,0.510)$~GeV/$c^2$ are vetoed to reduce the
$K_S^0 \to \pi^+\pi^-$ background by 86\%.

For the reconstruction mode of $\EtapEta$, to increase the detection
efficiency, the neutral decays of $\eta$ are not reconstructed with
individual final states, but considered as missing.  Thus,
the $\eta$ signal is selected in the system recoiling against the ST
$\LamCB$, $p$, $\pi^+$ and $\pi^-$. To improve the mass resolution, we
use $M_{{\rm rec} (\RecoilEta)}+M_{\LamCB}-m_{\LamC}$ instead of
$M_{{\rm rec} (\RecoilEta)}$ to reconstruct the $\eta$ signal, where
$M_{{\rm rec} (\RecoilEta)}$ is the recoiling mass against the ST
$\LamCB$, $p$, $\pi^{+}$ and $\pi^{-}$, $M_{\LamCB}$ is the
reconstructed mass of the $\LamCB$ candidate, and $m_{\LamC}$ is the
world average $\LamC$ mass~\cite{PDG:2020}.  The $M_{{\rm rec}
(\RecoilEta)}$ variable is required to be within the $\eta$ mass interval
$(0.533, 0.561)$~GeV/$c^2$, corresponding to $1\sigma$ resolution of the
recoiling mass, to reject the $\LamC \to pK_{L}^{0} \pi^+
\pi^-$ background. The recoiling mass against the ST $\LamCB$ is required
to be within $(2.275, 2.310)$ GeV$/c^2$. Finally, the signal yield of
$\LamCPE$ is determined by fitting the distribution of $M_{{\rm rec}
(\LamCB p)}$, which is defined as the recoiling mass against the ST
$\LamCB$ and $p$, as presented in Fig.~\ref{fig:fit}(a).

For the reconstruction mode of $\EtapGam$, the $\gamma$ candidates are
selected from photons not assigned to ST $\pi^0$s, and the one with
the minimum value of $|\dE_{p\pi^+\pi^-\gamma}|$ is kept for further
analysis, where $\dE_{p\pi^+\pi^-\gamma}= E_{p\pi^+\pi^-\gamma} -
\Ebeam$ and $E_{p\pi^+\pi^-\gamma}$ is the reconstructed energy of the
candidate events. The candidate events are further required to be
within $\dE_{p\pi^+\pi^-\gamma} \in (-0.017, 0.008)$~GeV. The
background containing extra $\pi^{0}$s is vetoed with the requirement
of $M_{{\rm rec} (\RecoilEta)}<0.1$~GeV/$c^2$.  Additionally, the
$\gamma$ is selected with $\alpha_{\gamma}<20^{\degree}$, where
$\alpha_{\gamma}$ is the angle between $\gamma$ and the direction of
recoiling system of ST $\LamCB$, $p$, $\pi^{+}$ and $\pi^{-}$.
Furthermore, the invariant mass of $p\pi^+\pi^-\gamma$ is required to
be within $M_{p\pi^+\pi^-\gamma} \in (2.275, 2.310)$~GeV/$c^2$.
Finally, the signal yield of $\LamCPE$ is determined by fitting the
distribution of $M_{\pi^+\pi^-\gamma}$, as presented in
Fig.~\ref{fig:fit}(b).

\section{Background analysis}

\begin{figure}[!htp]
    \begin{center}
            \includegraphics[width=0.5\textwidth]{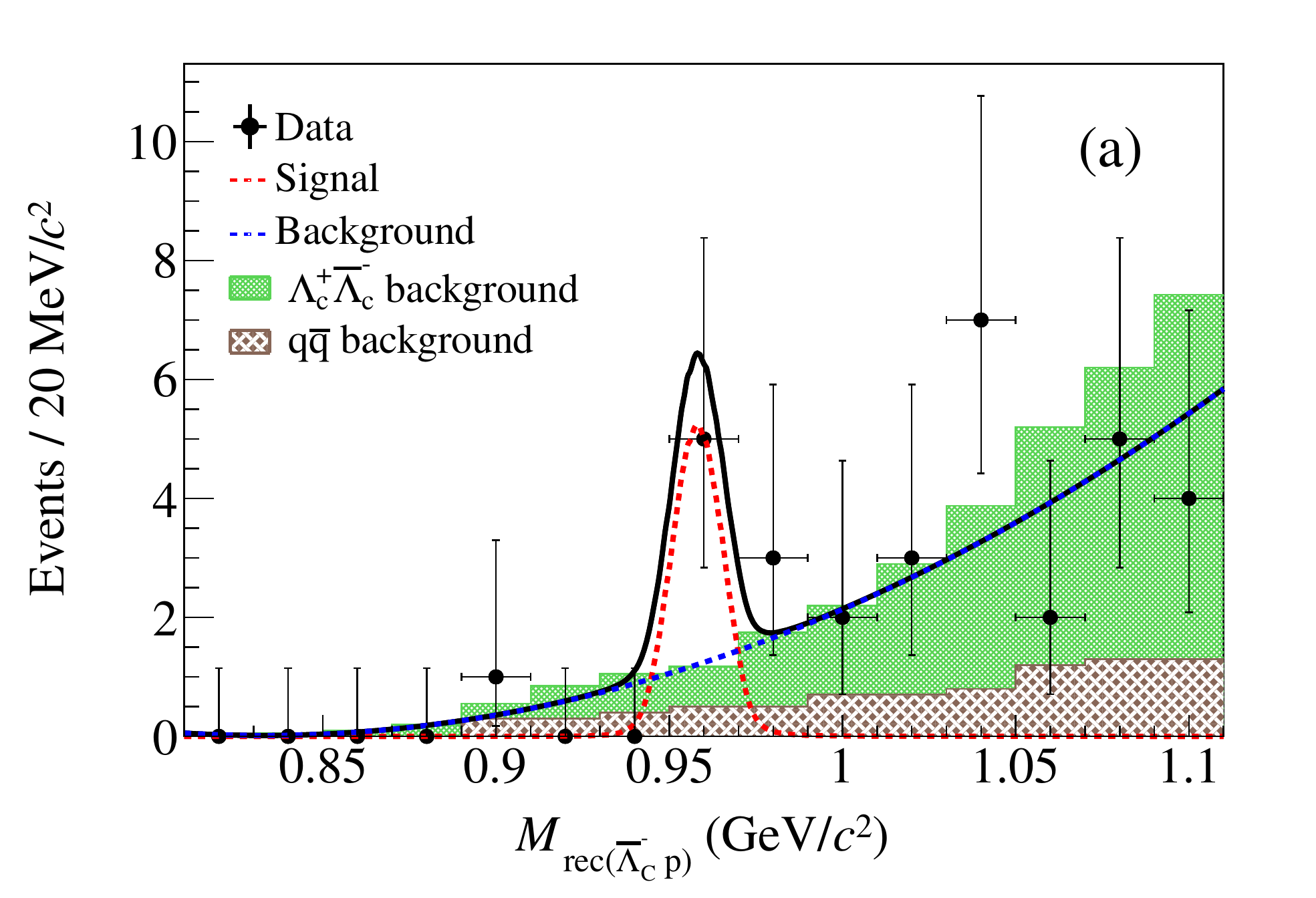}
            \includegraphics[width=0.5\textwidth]{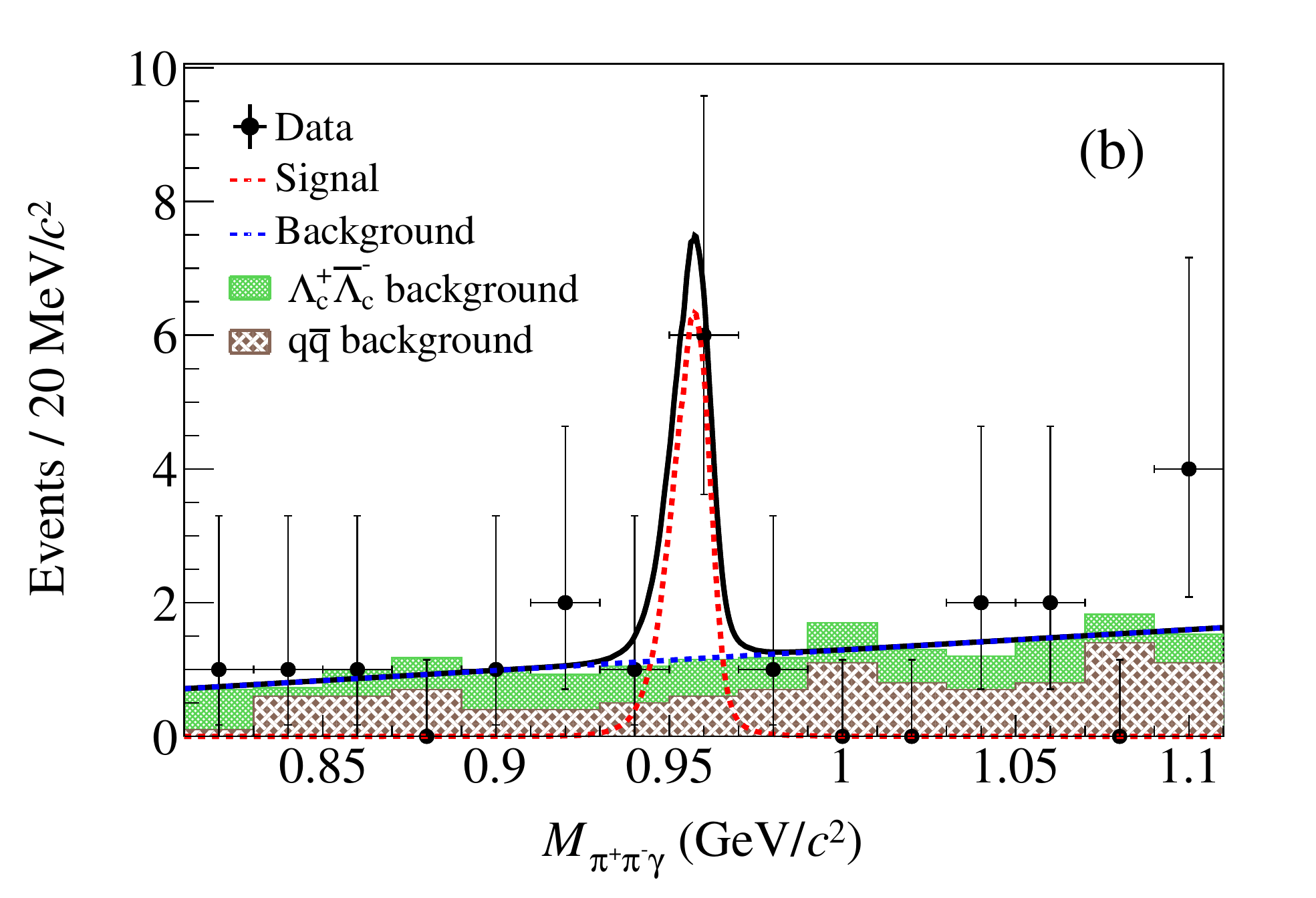}
    \end{center}
        \caption{ Simultaneous fit to (a) the $M_{{\rm rec} (\LamCB p)}$ distribution in the $\EtapEta$ mode and
              (b) the $M_{\pi^+\pi^-\gamma}$ distribution in the
              $\EtapGam$ mode with the combined seven data samples.
              The black points with error bars are data. The red and blue dashed lines indicate the curves for the signal and background, respectively.
              The black line is the sum over all the components in the fit.
              The brown and green hatched histograms for the two background components are from the inclusive MC samples.}
    \label{fig:fit}
\end{figure}

The potential background events are classified into two categories:
those directly originating from continuum hadron production in the
$\ee$ annihilation (denoted as $q\bar{q}$ background hereafter) and
those from the $\ee\to\LCpair$ events (denoted as $\LCpair$ background
hereafter). The distributions and magnitudes of $q\bar{q}$ and
$\LCpair$ backgrounds are estimated with the inclusive MC samples.
The main residual background sources are $\LamC \to
pK_{L}^0\pi^+\pi^-$, $\LamC \to \Sigma^+\eta$ and $\LamC \to
\Sigma^+\omega$.  The resultant $M_{{\rm rec} (\LamCB p)}$ and
$M_{\pi^+\pi^-\gamma}$ distributions of the accepted candidates in
data are depicted in Fig.~\ref{fig:fit}.  There are small peaks in the
$\eta^{\prime}$ signal regions for both decay modes. The simulated
shapes, which display no peaking background, describe the backgrounds
well.

\section{Branching Fraction Measurement}

The branching fraction of $\LamCPE$ is determined with
Eq.~(\ref{eq:br}) by performing a simultaneous unbinned maximum
likelihood fit on the distributions of $M_{\rm rec(\RecoilEtap)}$ and
$M_{\pi^{+}\pi^{-}\gamma}$ in the two $\eta^{\prime}$ decay modes,
constrained to the same $\mathcal{B}(\LamCPE)$ and taking into account
different detection efficiencies and branching fractions of
$\eta^{\prime}$.  The signal shapes are modeled by the MC-simulated
shapes. The background shapes are described by second-order polynomial
functions with fixed parameters, which are obtained by fitting the
corresponding distributions of inclusive MC samples, and the
background yields are floating.  The fit curves are depicted in
Fig.~\ref{fig:fit}. 
In Eq.~(\ref{eq:br}), the $\br_{\rm inter}$ values are (30.7$\pm$0.4)\% and (29.5$\pm$0.4)\% for the modes of $\EtapEta$ and $\EtapGam$, respectively, the ST detection efficiency $\epsilon_{ij}^{\mathrm{ST}}$ is obtained with the same procedure as
in Ref.~\cite{Ablikim:2022SCS}, and the DT detection efficiency
$\epsilon_{ij}^{\mathrm{DT}}$ is derived with the signal MC samples.
The efficiencies are summarized in
\tablename~\ref{tab:efficiency-DT-eta} and
\tablename~\ref{tab:efficiency-DT-gam} for the $\EtapEta$ and
$\pi^+\pi^-\gamma$ decay modes, respectively.  The statistical
significance of $\LamCPE$ is $3.6\sigma$, which is calculated from the
change of the likelihood values between the fits with and without the
signal component included, and accounting for the change in the number
of degrees-of-freedom.  The branching fraction of $\LamCPE$ is
determined to be $\mathcal{B}(\LamCPE) = (5.62^{+2.46}_{-2.04})\times
10^{-4}$, corresponding to signal yields of $4.9^{+3.2}_{-2.6}$ and
$4.3^{+2.6}_{-2.2}$ for the $\EtapEta$ and $\pi^+\pi^-\gamma$ modes,
where the uncertainties are statistical.

\begin{table}[!htbp]
  \begin{center}
  \caption{ The DT detection efficiencies in percentage in the
    $\EtapEta$ mode for the ten tag modes and seven data sets at different CM energies.
    The statistical uncertainties are lower than 0.3\%.}
    \begin{tabular}{ l r r r r r r r}
      \hline
      \hline
          Data set   &    $4600$   &  $4612$   &   $4628$   &    $4641$   &   $4661$   &   $4682$   &  $4699$  \\
      \hline

            $\Bpkpi$                       &  16.1  &  14.8  &  14.1 &  14.2  &  13.5  & 13.2  &  12.9 \\
            $\Bpks$                        &  18.6  &  15.6  &  15.7 &  14.8  &  14.5  & 14.1  &  14.3 \\
            $\bar{\Lambda}\pi^-$           &  15.2  &  12.8  &  12.3 &  11.7  &  11.7  & 10.9  &  10.4 \\
            $\Bpkpi\pi^0$                  &  3.9  &  3.7  &  3.4 &  3.3  &  3.3  & 3.2  &  3.2  \\
            $\Bpks\pi^0$                   &  5.3  &  4.5  &  4.6 &  4.2  &  4.5  & 3.8  &  4.0  \\
            $\bar{\Lambda}\pi^-\pi^0$      &  4.7  &  3.8  &  3.6 &  3.6  &  3.4  & 3.2  &  3.2  \\
            $\Bpks\pi^+\pi^-$              &  6.3  &  5.3  &  5.1 &  4.9  &  4.5  & 4.8  &  4.6  \\
            $\bar{\Lambda}\pi^-\pi^+\pi^-$ &  3.8  &  3.2  &  3.4 &  3.1  &  2.9  & 3.2  &  2.8  \\
            $\bar{\Sigma}^0\pi^-$          &  6.3  &  5.8  &  5.7 &  5.0  &  4.4  & 4.2  &  4.5  \\
            $\bar{\Sigma}^-\pi^+\pi^-$     &  5.3  &  4.7  &  4.7 &  4.3  &  4.1  & 3.9  &  4.0  \\
      \hline\hline
    \end{tabular}
    \label{tab:efficiency-DT-eta}
  \end{center}
\end{table}

\begin{table}[!htbp]
  \begin{center}
  \caption{ The DT detection efficiencies in percentage in the
    $\EtapGam$ mode for ten tag modes and seven data sets at different CM energies.
    The statistical uncertainties are lower than 0.3\%.}

    \begin{tabular}{ l r r r r r r r}
      \hline
      \hline
         Data set   &   $4600$   &  $4612$   &   $4628$   &    $4641$   &   $4661$   &   $4682$   &  $4699$  \\
      \hline

            $\Bpkpi$                       &  13.7  &  13.0   &  12.4   &  12.4   &  12.1   &  11.7  &  11.5 \\
            $\Bpks$                        &  15.4  &  14.3   &  14.0   &  14.6   &  13.4   &  12.7  &  12.5 \\
            $\bar{\Lambda}\pi^-$           &  13.2  &  11.0   &  10.4   &  10.6   &  10.8   &  9.9  &  9.8 \\
            $\Bpkpi\pi^0$                  &  3.7  &  3.5   &  3.4   &  3.3   &  3.3   &  3.2  &  3.2 \\
            $\Bpks\pi^0$                   &  4.2  &  4.3   &  4.5   &  4.3   &  3.8   &  4.1  &  4.0 \\
            $\bar{\Lambda}\pi^-\pi^0$      &  4.1  &  3.8   &  3.6   &  3.4   &  3.4   &  3.1  &  3.2 \\
            $\Bpks\pi^+\pi^-$              &  5.4  &  4.8   &  4.4   &  4.6   &  4.8   &  4.8  &  4.6 \\
            $\bar{\Lambda}\pi^-\pi^+\pi^-$ &  3.7  &  3.4   &  3.1   &  3.0   &  3.0   &  3.0  &  3.0 \\
            $\bar{\Sigma}^0\pi^-$          &  6.9  &  5.0   &  5.4   &  5.0   &  4.5   &  5.1  &  4.5 \\
            $\bar{\Sigma}^-\pi^+\pi^-$     &  4.6  &  4.4   &  4.3   &  3.8   &  3.8   &  3.8  &  3.9 \\

      \hline\hline
    \end{tabular}
    \label{tab:efficiency-DT-gam}
  \end{center}
\end{table}

\section{Systematic Uncertainty}

The systematic uncertainties for the branching fraction measurement
comprise those associated with the ST yields ($\Nsingle$), detection
efficiencies of the ST $\LamCB$ ($\effsingle$), detection
efficiencies of the DT events ($\effdouble$) and signal yield ($N_{\mathrm{sig}}$).  
As the DT technique is
adopted, the systematic uncertainties originating from reconstructing
the ST side largely cancel.  The systematic uncertainties are
evaluated relative to the measured branching fraction.  The details
are described in the following.

The uncertainties associated with the tracking and PID efficiencies for
the proton and $\pi$ are determined with the control samples $J/\psi \to
p\bar{p} \pi^+\pi^-$~\cite{Uncer:pidTracking} and
$J/\psi\to\pi^+\pi^-\pi^0$~\cite{Ablikim:2017systpi}, respectively.
The systematic uncertainties for tracking and PID are both assigned to be
1.0\% for each proton and $\pi$, respectively.

The uncertainty in the reconstruction efficiency for the $\gamma$ in the $\EtapGam$
decay mode is assigned to be 1.0\% based on a study with the control sample
$J/\psi \to \rho^0\pi^0$ events~\cite{Ablikim:gamerr}.

The uncertainty in the ST yields is 0.5\%, which arises from the
statistical uncertainty and the fits to the $\mBC$ distributions. The
uncertainty in the fitting procedure is evaluated by floating the
truncation parameter of the ARGUS function and changing the single
Gaussian function to a double Gaussian function.

The systematic uncertainty of the $M_{{\rm rec} (\RecoilEta)}$
requirement for the $\EtapEta$ decay mode is estimated by correcting
the variable $M_{{\rm rec} (\RecoilEta)}$ in the MC samples according
to the observed resolution difference between data and MC simulation.
The resolution difference is studied with the control sample
$\Lambda_{c}^+ \to p K_L^0\pi^+\pi^-$.  The change of the obtained
efficiency of the corrected MC samples from the nominal efficiency,
1.3\%, is taken as the corresponding systematic uncertainty.

The systematic uncertainty due to the $\dE_{p\pi^+\pi^-\gamma}$
requirement for the $\EtapGam$ decay mode is studied with the control
sample $\Lambda_{c}^+ \to \Sigma^{0}\pi^+$ with $\Sigma^{0} \to
\Lambda\gamma$, which has a similar final state as the signal process.
The difference between the efficiency in MC simulation and that in the
control sample, 1.1\%, is assigned to be the systematic uncertainty.

The systematic uncertainties arising from the $M_{{\rm rec}
  (\RecoilEta)}$ and $\alpha_{\gamma}$ requirements for the $\EtapGam$
decay mode are studied with the control sample $\Lambda_{c}^+ \to p
K_{S}^0\pi^{0}$.  The differences of the efficiencies between data and
MC simulation, 1.3\% and 0.8\%, are taken as the systematic
uncertainties due to the $M_{{\rm rec} (\RecoilEta)}$ and
$\alpha_{\gamma}$ requirements, respectively.

The uncertainty in the fit strategy of $M_{{\rm rec} (\LamCB p)}$ and
$M_{\pi^+\pi^-\gamma}$ is assigned to be 2.9\%, which is estimated by
shifting the fit range higher by 50 MeV and varying the background
shape to a third-order polynomial function.

The systematic uncertainties due to the signal shape for the
$\EtapEta$ and $\EtapGam$ decay modes are studied with the control
samples $\Lambda_{c}^+ \to p K_{S}^{0}$ and $\Lambda_{c}^+ \to
\Sigma^{0}\pi^+$ with $\Sigma^{0} \to \Lambda\gamma$, respectively.
The distributions of $M_{{\rm rec} (\LamCB p)}$ and $M_{\Sigma^{0}}$
are fitted by MC shapes, with or without a convolution with a
Gaussian function.  The differences of yields between the two cases,
0.6\% and 0.1\%, are taken as the corresponding systematic
uncertainties for the two modes, respectively.

The systematic uncertainty arising from the signal modeling is
investigated by generating a new set of signal MC $\LamCPE$ events, with
proton polar angle distribution parameterized by $1+\alpha {\rm
  cos}^2\theta$~\cite{Ablikim:alpha} with decay parameter $\alpha =
\pm 1$.  Here $\theta$ is the polar angle of the proton with respect
to the $\LamC$ in the $\ee$ CM system.  Comparing the differences in
detection efficiencies between the nominal and alternative samples,
the resultant uncertainties are obtained to be 1.0\% for the
$\EtapGam$ decay mode, and 0.1\% for the $\EtapEta$ mode.

The uncertainties in the branching fractions of the intermediate state
decays from the PDG~\cite{PDG:2020} are 0.5\% and 0.4\% for the
$\EtapEta$ and $\EtapGam$ decay modes, respectively.

According to Eq.~(\ref{eq:br}), the uncertainty related to the ST
efficiency mostly cancels. However, due to different
multiplicities, the ST efficiencies estimated with the generic and
signal MC samples are expected to differ from each other
slightly and result in a so called “tag bias”
uncertainty.  The difference between the ST efficiencies given by the
generic and signal MC samples, 0.9\%, is assigned as the corresponding
uncertainty.

Table~\ref{tab:sys-sum} summarizes the individual relative systematic uncertainties,
where the correlated systematic uncertainties are listed in the top and 
the uncorrelated systematic uncertainties are listed in the bottom.

\begin{table}[htbp]
  \begin{center}
  \caption{The relative systematic uncertainties in percent for $\LamCPE$ in the decay modes of $\EtapEta$ and $\EtapGam$.}\label{tab:sys-sum}
   \setlength{\tabcolsep}{2mm}{
    \begin{tabular}{ l c c }
      \hline
      \hline
          Source   &  $\pi^+\pi^-\eta$   &  $\pi^+\pi^-\gamma$  \\
      \hline
            $p$ tracking                                                           &  \multicolumn{2}{c}{1.0} \\
            $p$ PID                                                                &  \multicolumn{2}{c}{1.0} \\
            $\pi$ tracking                                                         &  \multicolumn{2}{c}{2.0} \\
            $\pi$ PID                                                              &  \multicolumn{2}{c}{2.0} \\
            ST yield                                                               &  \multicolumn{2}{c}{0.5} \\
            $M_{{\rm rec} (\RecoilEta)}$ requirement                               &  \multicolumn{2}{c}{1.3} \\
            Fit strategy                                                           &  \multicolumn{2}{c}{2.9} \\               
            Tag bias                                                               &  \multicolumn{2}{c}{0.9} \\
      \hline
            $\gamma$ detection                                                     & -            & 1.0 \\
            $\dE_{p\pi^+\pi^-\gamma}$ requirement                                  & -            & 1.1 \\
            $\alpha_{\gamma}$ requirement                                          & -            & 0.8 \\
            $\br_{\rm inter}$                                                      & 0.5          & 0.4 \\
            Signal shape                                                           & 0.6          & 0.1 \\ 
            Signal modeling                                                        & 0.1   & 1.0 \\ 
      \hline
            Total                                                                  & 4.7          & 5.0 \\
      \hline\hline
    \end{tabular}
  }
  \end{center}
\end{table}

\section{Summary}
\label{sec:summary}

In summary, the SCS decay $\LamCPE$ is observed with a statistical
significance of $3.6\sigma$ by using $\ee$ collision data samples
corresponding to a total integrated luminosity of 4.5~$\invfb$
collected at seven CM energies between 4.600 and 4.699~GeV with the BESIII
detector.  The absolute branching fraction of $\LamCPE$ is measured to
be $(5.62^{+2.46}_{-2.04} \pm 0.26)\times 10^{-4}$, where the first
and second uncertainties are statistical and systematic, respectively.
The branching fraction measured in this work is consistent with the
Belle result~\cite{Belle:2022etap} and the predictions in
Refs.~\cite{Sharma:1996sc, Geng:2018rse}, but significantly higher
than that in Ref.~\cite{Khanna:1994sc}, as shown in
\tablename~\ref{tab:BFs-sum}.  The result from this analysis provides
an input to understand the dynamics of charmed baryon
decays, and helps to improve different theoretical models.

\begin{table}[htbp]
  \begin{center}
  \caption{Comparison of the measured branching fraction (in $10^{-4}$) of $\LamCPE$ to theoretical predictions and the Belle result.}\label{tab:BFs-sum}
   \setlength{\tabcolsep}{2mm}{
    \begin{tabular}{ l c }
      \hline
      \hline
            &  $\LamCPE$  \\
      \hline
            BESIII                                           &            $5.62^{+2.46}_{-2.04} \pm 0.26$             \\
            Belle~\cite{Belle:2022etap}                      &            $4.73 \pm 0.97$                             \\
            Sharma {\it et al.}~\cite{Sharma:1996sc}         &            $4 - 6$                                       \\
            Uppal {\it et al.}~\cite{Khanna:1994sc}          &            $0.4 - 2$                                   \\
            Geng {\it et al.}~\cite{Geng:2018rse}            &            $12.2^{+14.3}_{-8.7}$                       \\
      \hline\hline
    \end{tabular}
  }
  \end{center}
\end{table}

\section*{ACKNOWLEDGMENT}
The BESIII collaboration thanks the staff of BEPCII and the IHEP computing center and the supercomputing center of
the University of Science and Technology of China (USTC) for their strong support. 
This work is supported in part by National Key R\&D Program of China under Contracts Nos.\ 2020YFA0406400, 2020YFA0406300; 
National Natural Science Foundation of China (NSFC) under Contracts Nos.\ 11635010, 11735014, 11835012, 11935015, 11935016, 
11935018, 11961141012, 12022510, 12025502, 12035009, 12035013, 12192260, 12192261, 12192262, 12192263, 12192264, 12192265, 12005311; 
the Fundamental Research Funds for the Central Universities, Sun Yat-sen University, University of Science and Technology of China;
100 Talents Program of Sun Yat-sen University;
the Chinese Academy of Sciences (CAS) Large-Scale Scientific Facility Program;
Joint Large-Scale Scientific Facility Funds of the NSFC and CAS under Contract No.\ U1832207; 
100 Talents Program of CAS; 
The Institute of Nuclear and Particle Physics (INPAC) and Shanghai Key Laboratory for Particle Physics and Cosmology; 
ERC under Contract No.\ 758462; 
European Union's Horizon 2020 research and innovation programme under Marie Sklodowska-Curie grant agreement under Contract No.\ 894790; 
German Research Foundation DFG under Contracts Nos.\ 443159800, Collaborative Research Center CRC 1044, GRK 2149; 
Istituto Nazionale di Fisica Nucleare, Italy; 
Ministry of Development of Turkey under Contract No.\ DPT2006K-120470; 
National Science and Technology fund; 
National Science Research and Innovation Fund (NSRF) via the Program Management Unit for Human Resources \& Institutional Development, 
Research and Innovation under Contract No.\ B16F640076; 
STFC (United Kingdom); 
Suranaree University of Technology (SUT), Thailand Science Research and Innovation (TSRI), 
and National Science Research and Innovation Fund (NSRF) under Contract No.\ 160355; 
The Royal Society, UK under Contracts Nos.\ DH140054, DH160214; 
The Swedish Research Council; 
U.S. Department of Energy under Contract No.\ DE-FG02-05ER41374.

\end{document}